\documentclass[aps,amsmath,amssymb,prl,a4paper,superscriptaddress,showpacs,twocolumn]{revtex4}
\usepackage{graphicx} %\usepackage{dcolumn} %\usepackage{bm}
\allowdisplaybreaks

\newcommand{\eqn}[1]{(\ref{#1})}
\newcommand{\beq}{\begin{equation}} 
\newcommand{\eneq}{\end{equation}}
\newcommand{\bea}{\begin{eqnarray}} 
\newcommand{\enea}{\end{eqnarray}}
\newcommand{\bean}{\begin{eqnarray*}}
\newcommand{\eean}{\end{eqnarray*}}

\newcommand{\dagga}{{\phantom{\dagger}}}
\newcommand{\fract}[2]{\frac{\displaystyle #1}{\displaystyle #2}}
\begin{document}

\title{Suppression of Kondo-assisted co-tunneling in a spin-1 quantum dot with Spin-Orbit interaction}

\author{Procolo Lucignano} 
\affiliation{CNR-SPIN, Monte S.Angelo -- via Cinthia,  I-80126 Napoli, Italy}
\affiliation{International School for Advanced Studies (SISSA/ISAS) 
Via Beirut 2-4, I-34151 Trieste, Italy} 
\author{Michele Fabrizio}
\affiliation{International School for Advanced Studies (SISSA/ISAS) 
Via Beirut 2-4, I-34151 Trieste, Italy}
\affiliation{CNR-IOM,Via Beirut 2-4, I-34151 Trieste, Italy} 
\affiliation{The Abdus Salam International Centre for Theoretical Physics 
(ICTP), P.O.Box 586, I-34151 Trieste, Italy}
\author{A. Tagliacozzo}
\affiliation{CNR-SPIN, Monte S.Angelo -- via Cinthia,  I-80126 Napoli, Italy}
\affiliation{Dipartimento di Scienze Fisiche, Universit\`a di
Napoli ``Federico II'', Monte S.Angelo -- via Cinthia, I-80126 Napoli, Italy}

\begin{abstract} Kondo-type zero-bias anomalies have been frequently observed in quantum dots 
occupied by two electrons and attributed to a spin-triplet
configuration that may become stable under particular circumstances. 
Conversely, zero-bias anomalies have been so far quite elusive
when quantum dots are occupied by an even number of electrons greater than two, even though a
spin-triplet configuration is more likely to be stabilized there than
for two electrons. We propose as an origin of this phenomenon the
spin-orbit interaction, and we show how it profoundly alters the conventional Kondo
screening scenario in the simple case of a laterally confined quantum dot with four electrons.
\end{abstract}

%\begin{keywords} Quantum Dots, Kondo effect, Numerical
%Renormalization Group, Spin orbit coupling \end{keywords}

\maketitle

%\section{Introduction}

%Since the first theoretical proposals, overwhelming evidences have been collected 
%that semiconductor Quantum Dots (QD) attached to metallic leads can conduct via Kondo-assisted 
%co-tunneling even when the electron number is locked by Coulomb blockade.\cite{}

A QD in the Coulomb blockade regime with an odd number of electrons acts as a localized magnetic moment 
and the spin degeneracy allows for Kondo effect to take 
place\cite{Glazman1,Ng&Lee,kastner,kouwenhoven2,aleiner} . Conversely, 
a QD with an even number of  electrons is usually in a non-degenerate spin-singlet configuration, 
hence the absence of any Kondo effect. 
For long time, the most direct signature of Kondo resonant tunneling was the so-called even-odd effect. 
In reality, the even-odd rule not always applies since also dots with an even number of electrons 
can become Kondo-active under an external field able to push a high spin 
configuration below the spin-singlet one.\cite{schmid,evenfield,Zumbuhl,pustilnik,arturo,kogan,2stage-exp}
This level crossing is called singlet-triplet (S-T) transition, since the high-spin state 
is usually a triplet, and is accompanied by several 
interesting phenomena.
\cite{2-stage_silvano,even,eto,pustilnik2,pustilnik3,2stage-dot,pustilnik4,even,hofstetter,kogan,2stage-exp} 
However the report of Kondo-like zero-bias anomalies in four or more electron dots \cite{schmid,2stage-exp} 
is rare when  compared with the wealth of data available for two-electron dots. In addition, even when 
these anomalies are indeed observed, like in the experiment by Granger {\sl et al.},\cite{2stage-exp} 
they are found to behave rather unconventionally as a function of temperature or magnetic field.

In this paper we propose that spin orbit interaction (SOI) may offer a natural explanation of the lack of 
Kondo-assisted co-tunneling in quantum dots with even number of electrons. 
When the number of electrons trapped in a QD increases, the  
separation between the single-particle orbitals lying closest to the chemical potential 
must diminish and eventually can be overwhelmed by the exchange splitting, thus stabilizing a magnetic 
ground state.\cite{procolo2,procolo} This is certainly the case for an axially symmetric dot 
with four electrons. It is well known that 
SOI may affect significantly magnetic properties of 
QD's,\cite{khaetskii_nazarov} a feature that attracts great interest 
in the context of quantum computation through 
semiconducting dots.\cite{SOI-qubit-1,SOI-qubit-2,SOI-qubit-3, SOI-two-electrons} 
By contrast, SOI is often not accounted for when interpreting tunneling spectra 
across quantum dots. On the contrary, we will show that SOI  
may actually affect dramatically quantum dots with an integer spin, especially 
when coupled only to a single conducting channel from the leads.     
In particular, we shall consider a very simplified model of a four-electron laterally confined dot 
and show by Numerical Renormalization Group (NRG)\cite{bullaRMP} that the SOI totally suppresses 
zero-bias conductance even when the four-electron ground state is a spin triplet. 
We will show that the zero-bias conductance has a non monotonic behavior in temperature 
and magnetic field, which strongly resembles the experimental data of 
Ref.\cite{2stage-exp} 

In a parabolic confining potential and in the absence of magnetic field,
the single particle eigenstates are those of a two dimensional harmonic oscillator with 
eigenvalues $\epsilon_{j} = \hbar \omega_0 (n_x+n_y+1)$, where $\omega_0$ is the
confinement frequency and $j=(n_x, n_y)$ labels the  states in a
cartesian basis. Exact diagonalization calculations show that, in the case of four electrons, the 
largest weight configuration in the ground state has two electrons filling the lowest-lying level, 
$j=(0,0)$, while the other two occupy the higher states, $j=(1,0)\equiv a$ and $(0,1)\equiv b$,  
in a spin-triplet cofiguration.\cite{procolo3} Therefore it is justified to consider the 
Hamiltonian of the isolated dot by including only the interaction within the $j=a,~b$ shell
\beq
H_{dot} =  \sum_{\sigma,j}\, \epsilon_j\,d^\dagger_{j\sigma}d^\dagga_{j\sigma}+  
U\sum_{j=a,b}\,n_{j\uparrow}n_{j\downarrow} - \fract{J_H}{2} \mathbf{S}\cdot\mathbf{S} \: ,
\label{dotto}
\eneq
where  $d^\dagger_{j\sigma}$ ($d^\dagga_{j \sigma}$) are the fermionic creation (annihilation) 
operators on the QD, respectively, and $n_{j\sigma}=d^\dagger_{j\sigma}d^\dagga_{j\sigma}$. 
Here  $\mathbf{S}$ is the total spin of the $a$-$b$ shell and  the Hund's 
term with $J_H\geq 0$ favors the triplet state.  
Spin-orbit coupling $H_{SO}$, involves, however, also the lowest lying levels $j=(0,0)$.  For a quantum dot 
defined in a two-dimensional electron layer, the SOI terms linear in momentum  
$\boldsymbol{\pi}=\mathbf{p}+e\mathbf{A}/c $  ($\mathbf{A}$ is the vector potential) are dominant, provided 
the dot lateral size is much larger than the layer thickness. We shall parametrize the 
Rashba term\cite{rashba} by $\alpha $, and the Dresselhaus\cite{dresselhaus} one by $\beta$ 
(ranging from tens to few hundreds of  $meV \cdot $\AA ). Thus
 \beq
 H_{SO} =-\fract{\alpha}{\hbar }\bigg(\pi_x\sigma_y-\pi_y \sigma_x 
\bigg) \:  -\fract{\beta}{\hbar}\bigg(\pi_x\sigma_x-\pi_y
\sigma_y \bigg) \: ,
%\label{spino} 
\nonumber
\eneq
where $\sigma_{x,y,z} $ are Pauli matrices.  As the typical energy scale of the 
SO coupling is much smaller than the bare single-particle level spacing $\hbar\omega_0$ 
($\alpha /\hbar$ and $\beta/\hbar$ $<< \sqrt{ \hbar \omega _0 /m_*} $),  it is legitimate  to treat $H_{SO}$ 
perturbatively. This amounts to degenerate second-order perturbation theory in $H_{SO}$ through intermediate 
excitated states with holes in the $j=(n_x,n_y)=(0,0)$ shell and/or electrons in the empty shell with 
$n_x+n_y=2$. The calculation is straightforward (see Ref.~\cite{Lucignano_physica_E} for details) and 
leads to 
\bea 
H_{SO}^{(2)}= -\lambda _s \sum_{\sigma ,
j=a,b} d^\dagger_{j\sigma}d^\dagga_{j\sigma}
 + i
\lambda \: \sum_{\sigma , j\neq j' }\sigma \:
d^\dagger_{j\sigma} d^\dagga_{j' \sigma} \:\: ,
\label{sorba}
\enea
with $\lambda _s =   m_*\left(\alpha ^2 + \beta ^2 \right) / \hbar ^2 $ and  
$\lambda  =  m_*\left(\alpha ^2 - \beta ^2 \right)  / \hbar ^2 $. Here, $m_*$  
is the effective mass of the semiconducting two dimensional layer (e.g. GaAs or InAs).  
In the presence of a magnetic field, parametrized in what follows by the cyclotron frequency $\omega_c$, 
$\epsilon _j $ as well as $\alpha$ and $\beta$ 
become field-dependent and a Zeeman splitting must be added to $H_{dot}$, Eq.(\ref{dotto}). 
The first term in Eq. (\ref{sorba}) shifts the position of $a,b$ with respect to 
the chemical potential,  breaking particle-hole symmetry and can always be 
compensated by changing the gate voltage. 
The second term of Eq. (\ref{sorba}) represents a spin dependent hopping between the two levels. 
Unlike the former, the latter it plays an important role 
that is more transparent at large $U$,  
as it provides an additional anisotropic contribution to the spin exchange 
besides the isotropic one $\propto J_H$. In this limit and at zero magnetic field, $H_{dot}$ and $H_{SO}$ 
can be mapped onto a simple spin-1 Hamiltonian:
\bea
H_{dot} +H_{SO} \longrightarrow  
-\frac{1}{2}\left(J_H+J_{SO}\right)\,\mathbf{S}\cdot\mathbf{S} + J_{SO}\,\left(S^z\right)^2 
\, .
\label{Hund+SO} 
\enea
Here $J_{SO}=4\lambda^2/U$.  The SOI thus generates a hard-axis single-ion anisotropy, 
which splits the spin-triplet into a lower state with $S^z=0$ and a higher doubly-degenerate one 
with $S^z=\pm 1$. It follows that SOI competes against Kondo effect, which instead requires a QD degenerate 
ground state. We shall see that, in the specific geometry we consider, this competition is 
actually won by SOI.

We now supplement the Hamiltonian $H_{dot}+ H_{SO}^{(2)} $ of Eqs.(\ref{dotto},\ref{sorba}) 
with the Hamiltonian of the leads $H_{lead}$, assumed to be free, and a 
term $H_{hyb}=\sum_{\sigma k } V_{k} \left (c^\dagger_{k\sigma} d^\dagga_{a\sigma} +H.c. \right )$ 
describing the hybridization to a suitable combination of  states $|k\sigma>$ from the two conducting leads. 
For sake of simplicity, we shall assume that electrons from the leads can tunnel only into one level, e.g. $a$.
Since Eq. \eqn{Hund+SO} is invariant under any rotation in the space of the two 
orbitals $a$ and $b$, the single screening channel could be coupled to a combination of 
both orbitals rather than to a single one, with no change of the physics.
Our model hamiltonian is very similar to the two impurity single channel Kondo model studied in 
Refs.\cite{refereea1,refereea2}. However, in our case the two levels are coupled 
ferromagnetically and the interesting physics arises by the SOI rather than by an antiferromagnetic 
exchange between the impurities as in \cite{refereea1,refereea2}. 

%We denote as $H_{lead}$ the free Hamiltonian of the leads. $H_{lead}+H_{hyb}$ define 
%a hybridization parameter $\Gamma$ that we shall assume finite and constant in energy up to a 
%cutoff $D$ of the order of the conduction bandwidth.  
%The full Hamiltonian thus reads $H=H_{dot}+ H_{SO}^{(2)} + H_{hyb}+H_{lead} $.  
%We will assume a flat density of states $\rho_o$ up to a cutoff  $D$ for the particle hole symmetric model 
%in the absence of SOI ($\alpha =\beta = 0$). 

According to Eq.\eqref{Hund+SO}, the physics of the model Hamiltonian $H$ for large $U$ is controlled 
by three energy scales: the Kondo screening temperature $T_{1\,K}$ of the level $a$ in the absence 
of any coupling to $b$, i.e. for $J_H=J_{SO}=0$, the Coulomb exchange, $J_{H}$, and finally 
the spin-orbit anisotropic exchange, $J_{SO}$. When $J_{SO}\gg T_{1\,K}$, the spin degeneracy 
is lost much before Kondo effect could start playing any role and the conductance must be 
small and structureless at low bias. A richer behavior instead emerges in the opposite limit 
of $J_{SO}\ll T_{1\,K}$. 
%%%%%%%%%%%
Here we can adopt a two-cutoff scaling approach and imagine to initially follow the system from 
high temperature/energy ($\gg J_{SO}$) as if SOI is absent. When the temperature/energy becomes of order 
$J_{SO}$, SOI  fully comes  into play.   
%%%%%%%%%%%
In this approximate scheme, on a scale $T_{1\,K}$ a first underscreened Kondo effect sets in, where   
only half of the dot-spin gets screened by the single conducting channel.\cite{Hewson-underscreened} 
The quasiparticles\cite{Nozieres-JLTP} that are coupled to the 
residual spin-1/2 acquire a local density-of-states (DOS) $\sim 1/T_{1\,K}$. 
The spin-1/2 that is left aside has a 
weak residual ferromagnetic exchange with the conduction bath whose effective strength $-J_*<0$  
vanishes at low temperature/energy.\cite{Hewson-underscreened}  
At an energy scale $\sim J_{SO}$, SOI modifies the 
effective exchange with the conduction bath into a spin-anisotropic one, see Eq.\eqref{Hund+SO}, with 
the coupling in the $x$-$y$ plane, $\sim -J_* - J_{SO}$, being larger in magnitude 
than that along $z$, $\sim -J_* + J_{SO}$. This case is known to lead to a further Kondo 
effect controlled by the Kondo temperature\cite{Hewson} 
\beq
T_{2\,K} \sim  T_{1\,K}\,\exp\left[-\fract{T_{1\,K}}{A}\left(\frac{\pi}{2}-\tan^{-1}\fract{J_z}{A}
\right)\right],\label{T2K}
\eneq    
where $A=2\,\sqrt{J_* J_{SO}}$. This looks like if a kind of two-stage Kondo effect takes place with 
well separated energy scales $T_{1\,K}\gg T_{2\,K}$, whose  low temperature phase is strongly 
driven by the spin dependent hopping due to the  SOI. The resemblance with recent findings 
on the role of magnetic anisotropies in models for magnetic 
impurities on surfaces \cite{Rock1,Rock2} (where single-ion anisotropies like 
in Eq.~\eqn{Hund+SO} emerge as well) is striking.
 
The above expectations that we drew from very qualitative arguments are nicely confirmed 
by the full NRG calculation.

 \begin{figure}[htbp]
\begin{center}
\includegraphics*[width=1\linewidth]{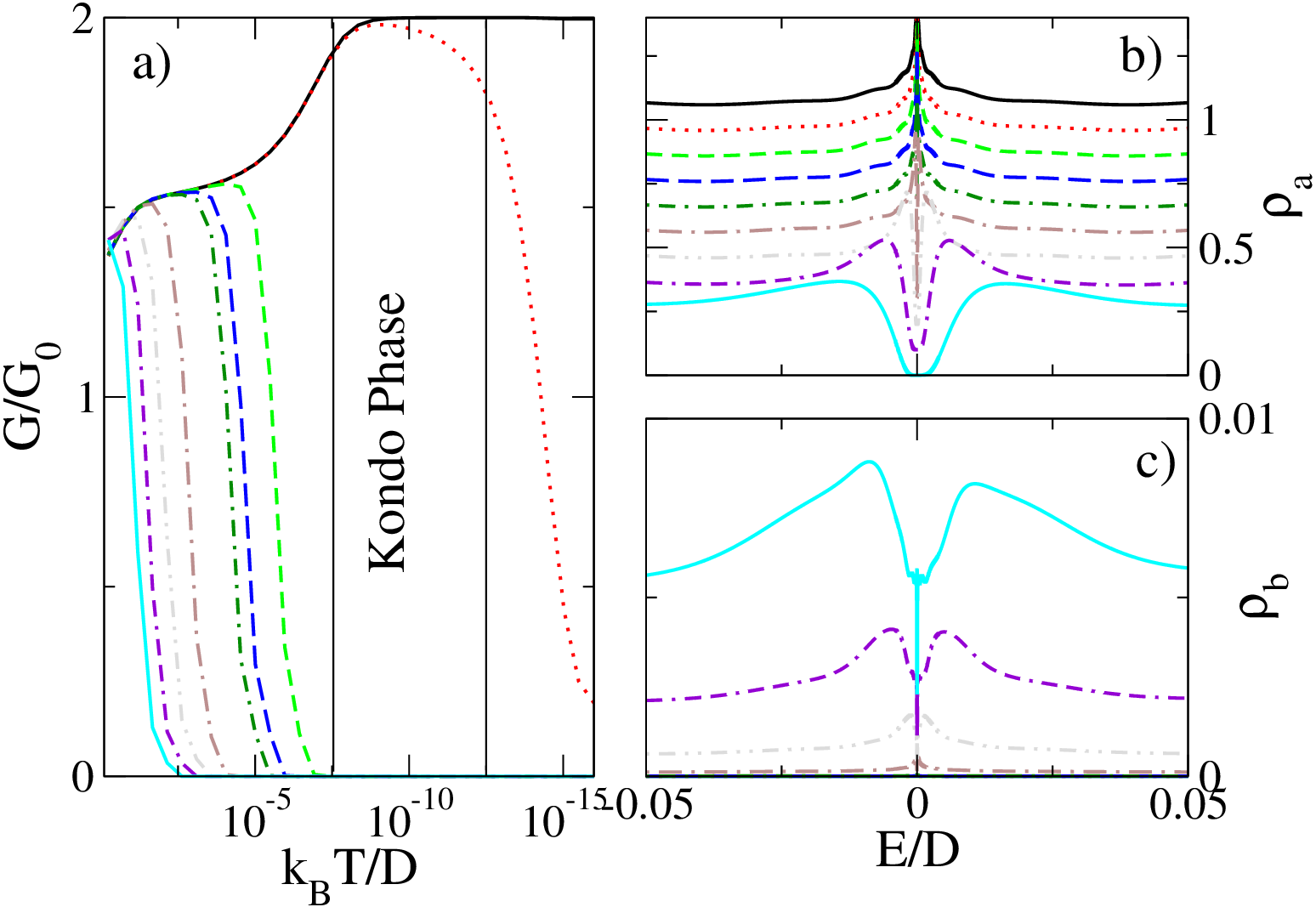}
\caption{(Color online) a): Zero bias conductance as a function of temperature 
for different values of $\lambda/\hbar\omega_0=0.0001,0.0025 ,0.0064 ,0.01,0.04 ,0.09, 0.16, 0.25$, 
increasing from the rightmost curve towards the leftmost one.
Hamiltonian parameters are: $\epsilon _a=\epsilon_b=-1meV $, $U =2meV$, 
$\Gamma= \pi \rho_0 |V_k|^2=0.1meV$, $D=1meV$, $J_H=0.1 meV$, $\omega_c=0$ and $\lambda_s=0$. 
The reference value $G_0=e^2/h$. 
b): Spectral function at the impurity site $a$. 
By increasing the spin orbit coupling the Kondo peak is suppressed and the central 
resonance turns into an antiresonance.  Curves have been shifted from clarity by a uniform amount and the scale on the y axis refers to the bottom curve.  From bottom to top $\lambda$ decreases.
c): Spectral function at the impurity site $b$ non directly connected to the contact leads.}
\label{GvsSO}
\end{center}
\end{figure}
The zero-bias conductance $G$ as function of the temperature is shown in Fig.~\ref{GvsSO}a 
for different $\lambda$'s. 
At very low temperatures $T\ll T_{2\,K}$, $G$ is indeed extremely small, practically zero. 
However, at intermediate temperatures $T_{2\,K}\ll T \ll T_{1\, K}$, the conductance can reach 
(or be very close to) its unitary value of an underscreened Kondo-like plateau.
The local spectral properties are shown in Fig.~\ref{GvsSO}b for the same values of $\lambda$ 
of Fig.~\ref{GvsSO}a.  
The density of states (DOS) $\rho_{a}(\epsilon)$ of the level $a$ that is coupled to the leads develops 
a conventional Abrikosov-Suhl resonance of width $T_{1\,K}$, 
signaling the underscreened Kondo effect. At lower energies,  a deep pseudogap of width $T_{2\,K}$ is digged inside the former resonance. 
Conversely, the DOS $\rho_{b}(\epsilon)$ of the level $b$, which hosts most of the 
residual spin-1/2, is quite low on the scale $\sim T_{1\,K}$ (note that it is two orders of magnitude smaller 
than $\rho_{a}$ in the figure) and develops a narrow antiresonance 
below $T_{2\,K}$. Since the zero-bias conductance $G$ is proportional to $\rho_{a}(0)$, 
$G$ has an inverted zero-bias anomaly, being small at zero temperature and increasing on increasing $T$.  
As long as only a single channel of conduction electrons is coupled to the dot, 
this behavior must hold whatever is the value of $\alpha^2-\beta^2\not = 0$, even if 
the space symmetry of the device is weakly perturbed. 
In fact, the ultimate cause of the ineffectiveness of the Kondo co-tunneling can be traced 
back to the spin-exchange \eqn{Hund+SO}, which is invariant under any unitary trasformation 
in the $a$-$b$ space. 
%%%%%%%%%%%%%%%%%
We note that, if the two orbitals are split or hybridized among each other because 
of an asymmetric shape of the confining potential, which is more likely the rule,   
the situation would not change provided the splitting and/or hybridization are small 
enough compared with $J_H$, so that the lowest energy state has still $S=1$.
However, should the splitting and/or hybridization be so large to stabilize a 
spin-singlet state of the dot, still Kondo co-tunneling would be ineffective.\cite{refereea1,refereea2}
Therefore, in the most general case of non-degenerate levels, we predict that, whatever is the magnitude 
of the Coulomb exchange $J_H$, Kondo-like zero-bias anomalies in four electron dots 
should be absent at low temperatures, provided a single conducting channel tunnels into the dot. 
If $J_H$ is small, this occurs because the dot electrons 
prefer to lock into a Kondo-inactive spin-singlet configuration.\cite{refereea1,refereea2} If $J_H$ is large, 
it is the unavoidably present SOI that stabilizes a non-degenerate state, 
i.e. the $S_z=0$ componenent of the spin-triplet.    
%%%%%%%%%%%%%%%%

The suppression of the Kondo effect due to the SOI is quite different from  
that caused by a magnetic field.
The magnetic field affects the whole low-energy ($\leq T_{1\,K}$) 
spectrum\cite{Hewson}; it splits the Abrikosov-Shul resonance and leads to a 
tiny zero-bias conductance that keeps decreasing on increasing temperature, see e.g. 
Refs.~\cite{Costi-magnetic-field,Costi-magnetic-field-underscreened}. 
By contrast, the SO coupling is gentler on the high energy scales $\sim T_{1\,K}$, but 
much more dramatic at low energy $\leq T_{2\,K}$. The 
Abrikosov-Shul resonance develops as usual, but in the end, the SOI  digs a narrow but very deep pseudogap 
at the chemical potential. Thus the conductance shows a Kondo plateau at intermediate 
temperatures, unlike what happens in the presence of a magnetic field, but falls down rapidly 
below $T_{2\,K}$ to much lower values than those at finite magnetic field. 

\begin{figure}[htbp]
\begin{center}
\includegraphics*[width=1\linewidth]{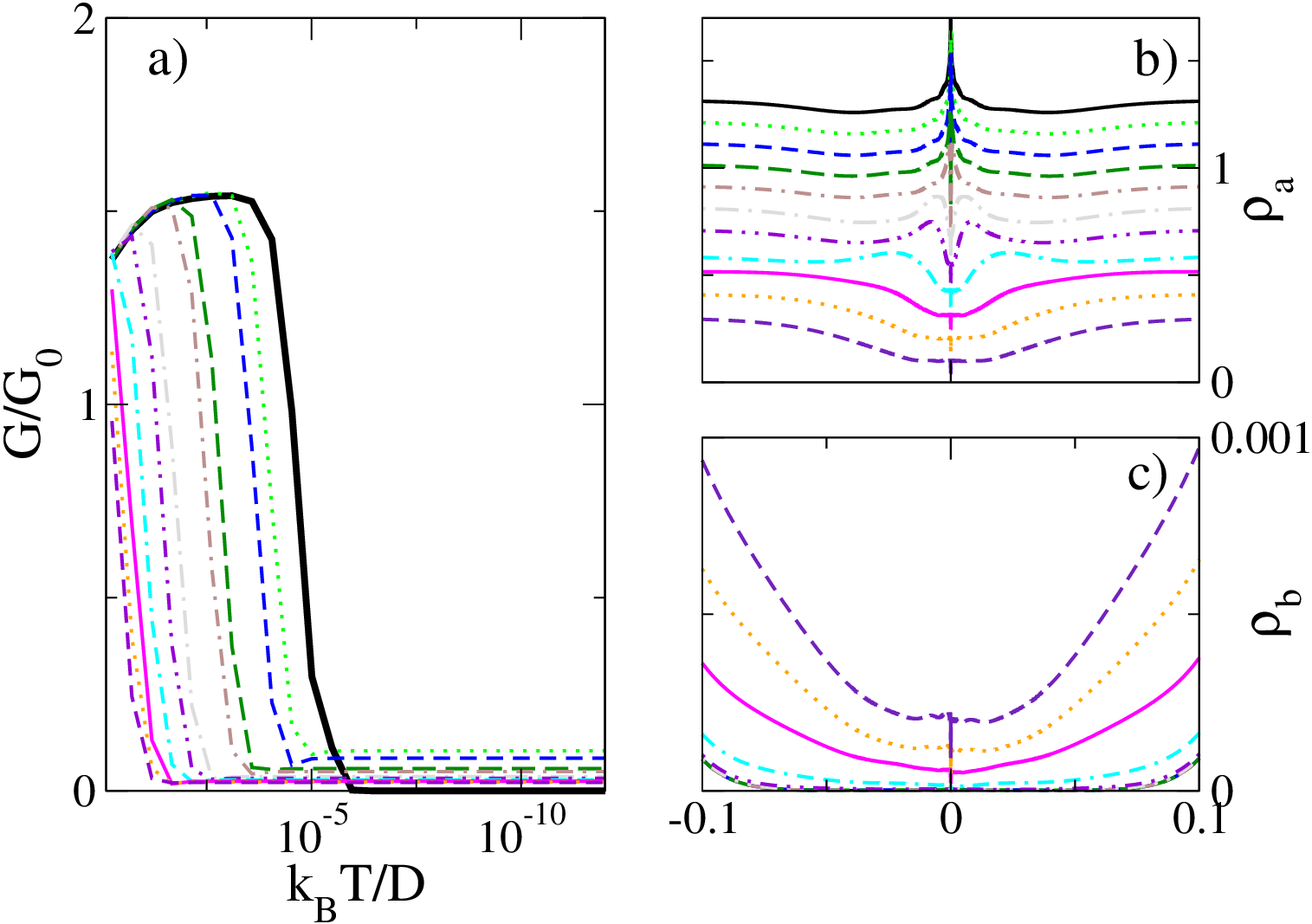}
\caption{(Color online) a): Zero bias conductance as a function of temperature for different values 
of $\omega_c/\omega_0=0.00005,0.0001,0.001,0.005,0.01,0.025,0.05,0.075,0.1$ (from right to left),  
where $\omega_c$ is the cyclotron frequency, at fixed spin orbit coupling $\lambda/\hbar\omega_0=0.01$. 
All other parameters are the same as in Fig.~\ref{GvsSO}. b): Spectral function of the $a$-level.
We note that by increasing the magnetic field the Kondo peak is suppressed and the central resonance 
turns into a wide antiresonance. Curves have been shifted from clarity by a uniform amount and the scale on the y axis refers to the bottom curve. 
From bottom to top $\omega_c/\omega_0$ decreases. c): Spectral function of the $b$-level.}
\label{GvsB}
\end{center}
\end{figure}

\begin{figure}[htbp]
\begin{center}
\includegraphics*[width=1\linewidth]{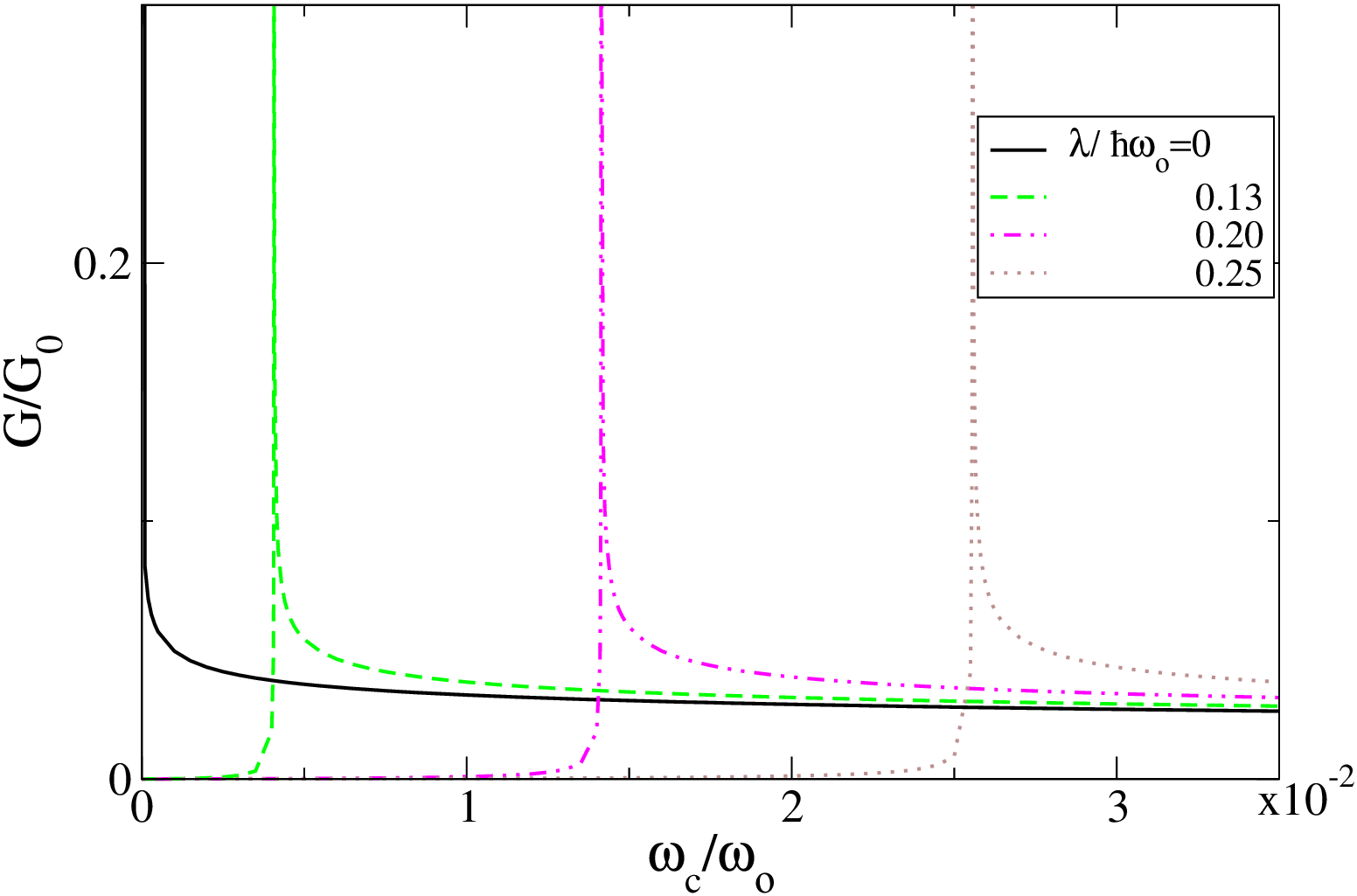}
\caption{(Color online)   Zero bias conductance at zero temperature 
as a function of the magnetic field $\omega_c/\omega_0$ 
for increasing values of the SOI $\lambda/\hbar \omega_0=0.0, 0.12, 0.2, 0.25$ 
(from left to right). }
\label{GzerovsB}
\end{center}
\end{figure}
%Annoto qui i coefficienti di scaling per nostra memoria
%0.3   0.8
%0.35 0.7 
%0.4   0.65
%0.45 0.58
%0.5   0.48
The combined action of  SOI and magnetic field is presented in   
Figs.~\ref{GvsB} and \ref{GzerovsB}. The intermediate underscreened Kondo phase  
disappears, no matter how small the magnetic field is  (see Fig.~\ref{GvsB}a).  
%%%%%%%%%%% NUOVA FRASE
Very weak magnetic fields give rise to a sudden drop of the conductance. 
In the absence of magnetic field the same result could be achieved only 
by means of unphysically large SOI. 
%%%%%%%%%%%%
The Kondo peak splits and a wide gap opens in the whole low energy region 
$\sim T_{1\,K}$ (see Fig.~\ref{GvsB}b), very similar to what found 
in the absence of SOI.\cite{Costi-magnetic-field,Costi-magnetic-field-underscreened}
The magnetoconductance is shown in Fig.~\ref{GzerovsB}  for increasing values of $\lambda$'s. 
The conductance first rises to a maximum at $\omega_c=\omega_c^*$ 
and then drops for large fields as $\sim 1/\omega_c^2$.% is (see Fig.~\ref{GzerovsB}b).   
By increasing $\lambda$ also  $\omega_c^*$ increases.
The sharp rise of the conductance at $\omega_c=\omega_c^*$ 
is an artifact of our simplified model and likely it will be rounded off in real devices. 

The non-monotonic behavior of $G$ both in temperature and in magnetic field 
has been observed experimentally in a four electron QD by Granger {\sl et al.}.\cite{2stage-exp} 
The explanation given by the authors invoked the two-stage Kondo effect proposed in 
Refs.~\cite{pustilnik4} and \cite{2stage-dot}. 
In that scenario, it is assumed that both the symmetric and the antisymmetric 
combination of the tunneling channels of each lead is coupled to the spin $S=1$ of the dot, so that 
eventually this spin gets fully screened although on two different temperature scales. The 
zero-bias conductance $G=G_0\sin^2 \delta$, where $\delta$ is the difference between the phase shifts of  
the symmetric and antisymmetric combinations, vanishes in that case since 
both channels acquire a $\pi/2$ phase shift.
$G$ as function of magnetic field or temperature turns out to be non-monotonous just like  
in our model. In spite of this, the two-stage Kondo effect and our scenario are very different.
Indeed, in the two-stage  Kondo effect of Ref\cite{pustilnik4,2stage-dot} both levels will have a Kondo 
peak at the Fermi level,
larger in $a$ than in $b$, while we do not find any in $b$.
Since the zero-bias conductance behaves similarly in both scenarios, it could 
be worth exploiting the tunability of the SOI to get further experimental insights. 
In the presence of SOI, the lowest lying state above the $S^z=0$ component of the spin-triplet 
should be the $S^z=\pm 1$ doublet, followed at higher energy by the singlet, a feature that could 
be uncovered by a detailed analysis of the inelastic tunneling spectrum in the absence and presence of 
a magnetic field. 
%We finally remark that in a real dot  with an increased number of electrons, 
%coupling to one channel only, is most likely.

In conclusion,  zero-bias anomalies have been so far quite elusive when quantum dots  are occupied by an even number of electrons greater than two,  even though a spin-triplet configuration is more likely to be stabilized  here than for two electrons (e.g. at zero magnetic field). Here we have proposed that a  possible explanation of the suppression of the Kondo conductance in an even electron quantum dot can be traced back to the role of the SOI,  which has been often disregarded in interpreting experiments. We have shown that SOI, in an underscreened four-electron  dot hybridized with one single channel, gives rise to a conductance behavior in the presence of a magnetic field, very close to what has been recently observed experimentally.\cite{2stage-exp}

\begin{acknowledgments}
The authors acknowledge Silvano De Franceschi 
for his critical reading of the manuscript and his interesting comments and suggestions.
Part of the numerical calculations have been done on the "nanomat" cluster 
owned by CNR-SPIN - UoS Napoli and Universita' di Napoli "Federico II"  
Dipartimento di Scienze Fisiche, which we kindly acknowledge.
P.L. acknowledges financial support from MIDAS-Macroscopic Interference Devices for Atomic and
Solid State Physics: Quantum Control of Supercurrents. M.F. has been supported by a MIUR-PRIN award.  
\end{acknowledgments}
After the submission of the paper  we became aware of Ref.\cite{Parks2010} stressing the importance of SOI in molecular transport.
\bibliographystyle{apsrev}
%\bibliography{bibdabozza.bib}

\begin{thebibliography}{42}
\expandafter\ifx\csname natexlab\endcsname\relax\def\natexlab#1{#1}\fi
\expandafter\ifx\csname bibnamefont\endcsname\relax
  \def\bibnamefont#1{#1}\fi
\expandafter\ifx\csname bibfnamefont\endcsname\relax
  \def\bibfnamefont#1{#1}\fi
\expandafter\ifx\csname citenamefont\endcsname\relax
  \def\citenamefont#1{#1}\fi
\expandafter\ifx\csname url\endcsname\relax
  \def\url#1{\texttt{#1}}\fi
\expandafter\ifx\csname urlprefix\endcsname\relax\def\urlprefix{URL }\fi
\providecommand{\bibinfo}[2]{#2}
\providecommand{\eprint}[2][]{\url{#2}}

\bibitem[{\citenamefont{Glazmann and Raikh}(1988)}]{Glazman1}
\bibinfo{author}{\bibfnamefont{L.~I.} \bibnamefont{Glazmann}} \bibnamefont{and}
  \bibinfo{author}{\bibfnamefont{M.~E.} \bibnamefont{Raikh}},
  \bibinfo{journal}{JETP Lett.} \textbf{\bibinfo{volume}{47}},
  \bibinfo{pages}{352} (\bibinfo{year}{1988}).

\bibitem[{\citenamefont{Ng and Lee}(1988)}]{Ng&Lee}
\bibinfo{author}{\bibfnamefont{T.~K.} \bibnamefont{Ng}} \bibnamefont{and}
  \bibinfo{author}{\bibfnamefont{P.~A.} \bibnamefont{Lee}},
  \bibinfo{journal}{Phys. Rev. Lett.} \textbf{\bibinfo{volume}{61}},
  \bibinfo{pages}{1768} (\bibinfo{year}{1988}).

\bibitem[{\citenamefont{Goldhaber-Gordon
  et~al.}(1998)\citenamefont{Goldhaber-Gordon, Shtrikman, Mahalu, Abush-Magder,
  Meirav, and Kastner}}]{kastner}
\bibinfo{author}{\bibfnamefont{D.}~\bibnamefont{Goldhaber-Gordon}},
  \bibinfo{author}{\bibfnamefont{H.}~\bibnamefont{Shtrikman}},
  \bibinfo{author}{\bibfnamefont{D.}~\bibnamefont{Mahalu}},
  \bibinfo{author}{\bibfnamefont{D.}~\bibnamefont{Abush-Magder}},
  \bibinfo{author}{\bibfnamefont{U.}~\bibnamefont{Meirav}}, \bibnamefont{and}
  \bibinfo{author}{\bibfnamefont{M.~A.} \bibnamefont{Kastner}},
  \bibinfo{journal}{Nature} \textbf{\bibinfo{volume}{391}},
  \bibinfo{pages}{156} (\bibinfo{year}{1998}).

\bibitem[{\citenamefont{Cronenwett et~al.}(1998)\citenamefont{Cronenwett,
  Oosterkamp, and Kouwenhoven}}]{kouwenhoven2}
\bibinfo{author}{\bibfnamefont{S.~M.} \bibnamefont{Cronenwett}},
  \bibinfo{author}{\bibfnamefont{T.~H.} \bibnamefont{Oosterkamp}},
  \bibnamefont{and} \bibinfo{author}{\bibfnamefont{L.~P.}
  \bibnamefont{Kouwenhoven}}, \bibinfo{journal}{Science}
  \textbf{\bibinfo{volume}{281}}, \bibinfo{pages}{540} (\bibinfo{year}{1998}).

\bibitem[{\citenamefont{Aleiner et~al.}(2002)\citenamefont{Aleiner, Brouwer,
  and Glazman}}]{aleiner}
\bibinfo{author}{\bibfnamefont{I.~L.} \bibnamefont{Aleiner}},
  \bibinfo{author}{\bibfnamefont{P.~W.} \bibnamefont{Brouwer}},
  \bibnamefont{and} \bibinfo{author}{\bibfnamefont{L.~I.}
  \bibnamefont{Glazman}}, \bibinfo{journal}{Phys. Rep.}
  \textbf{\bibinfo{volume}{358}}, \bibinfo{pages}{309} (\bibinfo{year}{2002}).

\bibitem[{\citenamefont{Schmid et~al.}(2000)\citenamefont{Schmid, Weis, Eberl,
  and v.~Klitzing}}]{schmid}
\bibinfo{author}{\bibfnamefont{J.}~\bibnamefont{Schmid}},
  \bibinfo{author}{\bibfnamefont{J.}~\bibnamefont{Weis}},
  \bibinfo{author}{\bibfnamefont{K.}~\bibnamefont{Eberl}}, \bibnamefont{and}
  \bibinfo{author}{\bibfnamefont{K.}~\bibnamefont{v.~Klitzing}},
  \bibinfo{journal}{Phys. Rev. Lett.} \textbf{\bibinfo{volume}{84}},
  \bibinfo{pages}{5824} (\bibinfo{year}{2000}).

\bibitem[{\citenamefont{Kouwenhoven et~al.}(1997)\citenamefont{Kouwenhoven,
  Oosterkamp, Danoesastro, Eto, Austing, Honda, and Tarucha}}]{evenfield}
\bibinfo{author}{\bibfnamefont{L.~P.} \bibnamefont{Kouwenhoven}},
  \bibinfo{author}{\bibfnamefont{T.~H.} \bibnamefont{Oosterkamp}},
  \bibinfo{author}{\bibfnamefont{M.~W.~S.} \bibnamefont{Danoesastro}},
  \bibinfo{author}{\bibfnamefont{M.}~\bibnamefont{Eto}},
  \bibinfo{author}{\bibfnamefont{D.~G.} \bibnamefont{Austing}},
  \bibinfo{author}{\bibfnamefont{T.}~\bibnamefont{Honda}}, \bibnamefont{and}
  \bibinfo{author}{\bibfnamefont{S.}~\bibnamefont{Tarucha}},
  \bibinfo{journal}{Science} \textbf{\bibinfo{volume}{278}},
  \bibinfo{pages}{1788} (\bibinfo{year}{1997}).

\bibitem[{\citenamefont{Zumb\"uhl et~al.}(2004)\citenamefont{Zumb\"uhl, Marcus,
  Hanson, and Gossard}}]{Zumbuhl}
\bibinfo{author}{\bibfnamefont{D.~M.} \bibnamefont{Zumb\"uhl}},
  \bibinfo{author}{\bibfnamefont{C.~M.} \bibnamefont{Marcus}},
  \bibinfo{author}{\bibfnamefont{M.~P.} \bibnamefont{Hanson}},
  \bibnamefont{and} \bibinfo{author}{\bibfnamefont{A.~C.}
  \bibnamefont{Gossard}}, \bibinfo{journal}{Phys. Rev. Lett.}
  \textbf{\bibinfo{volume}{93}}, \bibinfo{pages}{256801}
  (\bibinfo{year}{2004}).

\bibitem[{\citenamefont{Pustilnik et~al.}(2000)\citenamefont{Pustilnik,
  Avishai, and Kikoin}}]{pustilnik}
\bibinfo{author}{\bibfnamefont{M.}~\bibnamefont{Pustilnik}},
  \bibinfo{author}{\bibfnamefont{Y.}~\bibnamefont{Avishai}}, \bibnamefont{and}
  \bibinfo{author}{\bibfnamefont{K.}~\bibnamefont{Kikoin}},
  \bibinfo{journal}{Phys. Rev. Lett.} \textbf{\bibinfo{volume}{84}},
  \bibinfo{pages}{1756} (\bibinfo{year}{2000}).

\bibitem[{\citenamefont{Giuliano and Tagliacozzo}(2000)}]{arturo}
\bibinfo{author}{\bibfnamefont{D.}~\bibnamefont{Giuliano}} \bibnamefont{and}
  \bibinfo{author}{\bibfnamefont{A.}~\bibnamefont{Tagliacozzo}},
  \bibinfo{journal}{Phys. Rev. Lett.} \textbf{\bibinfo{volume}{84}},
  \bibinfo{pages}{4677} (\bibinfo{year}{2000}).

\bibitem[{\citenamefont{Kogan et~al.}(2003)\citenamefont{Kogan, Granger,
  Kastner, Goldhaber-Gordon, and Shtrikman}}]{kogan}
\bibinfo{author}{\bibfnamefont{A.}~\bibnamefont{Kogan}},
  \bibinfo{author}{\bibfnamefont{G.}~\bibnamefont{Granger}},
  \bibinfo{author}{\bibfnamefont{M.~A.} \bibnamefont{Kastner}},
  \bibinfo{author}{\bibfnamefont{D.}~\bibnamefont{Goldhaber-Gordon}},
  \bibnamefont{and}
  \bibinfo{author}{\bibfnamefont{H.}~\bibnamefont{Shtrikman}},
  \bibinfo{journal}{Phys. Rev. B} \textbf{\bibinfo{volume}{67}},
  \bibinfo{pages}{113309} (\bibinfo{year}{2003}).

\bibitem[{\citenamefont{Granger et~al.}(2005)\citenamefont{Granger, Kastner,
  Radu, Hanson, and Gossard}}]{2stage-exp}
\bibinfo{author}{\bibfnamefont{G.}~\bibnamefont{Granger}},
  \bibinfo{author}{\bibfnamefont{M.~A.} \bibnamefont{Kastner}},
  \bibinfo{author}{\bibfnamefont{I.}~\bibnamefont{Radu}},
  \bibinfo{author}{\bibfnamefont{M.~P.} \bibnamefont{Hanson}},
  \bibnamefont{and} \bibinfo{author}{\bibfnamefont{A.~C.}
  \bibnamefont{Gossard}}, \bibinfo{journal}{Phys. Rev. B}
  \textbf{\bibinfo{volume}{72}}, \bibinfo{pages}{165309}
  (\bibinfo{year}{2005}).

\bibitem[{\citenamefont{van~der Wiel et~al.}(2002)\citenamefont{van~der Wiel,
  De~Franceschi, Elzerman, Tarucha, Kouwenhoven, Motohisa, Nakajima, and
  Fukui}}]{2-stage_silvano}
\bibinfo{author}{\bibfnamefont{W.~G.} \bibnamefont{van~der Wiel}},
  \bibinfo{author}{\bibfnamefont{S.}~\bibnamefont{De~Franceschi}},
  \bibinfo{author}{\bibfnamefont{J.~M.} \bibnamefont{Elzerman}},
  \bibinfo{author}{\bibfnamefont{S.}~\bibnamefont{Tarucha}},
  \bibinfo{author}{\bibfnamefont{L.~P.} \bibnamefont{Kouwenhoven}},
  \bibinfo{author}{\bibfnamefont{J.}~\bibnamefont{Motohisa}},
  \bibinfo{author}{\bibfnamefont{F.}~\bibnamefont{Nakajima}}, \bibnamefont{and}
  \bibinfo{author}{\bibfnamefont{T.}~\bibnamefont{Fukui}},
  \bibinfo{journal}{Phys. Rev. Lett.} \textbf{\bibinfo{volume}{88}},
  \bibinfo{pages}{126803} (\bibinfo{year}{2002}).

\bibitem[{\citenamefont{Sasaki et~al.}(2000)\citenamefont{Sasaki, Franceschi,
  Elzerman, van~der Wiel, Eto, Tarucha, and Kouwenhoven}}]{even}
\bibinfo{author}{\bibfnamefont{S.}~\bibnamefont{Sasaki}},
  \bibinfo{author}{\bibfnamefont{S.~D.} \bibnamefont{Franceschi}},
  \bibinfo{author}{\bibfnamefont{J.~M.} \bibnamefont{Elzerman}},
  \bibinfo{author}{\bibfnamefont{W.~G.} \bibnamefont{van~der Wiel}},
  \bibinfo{author}{\bibfnamefont{M.}~\bibnamefont{Eto}},
  \bibinfo{author}{\bibfnamefont{S.}~\bibnamefont{Tarucha}}, \bibnamefont{and}
  \bibinfo{author}{\bibfnamefont{L.~P.} \bibnamefont{Kouwenhoven}},
  \bibinfo{journal}{Nature} \textbf{\bibinfo{volume}{405}},
  \bibinfo{pages}{764} (\bibinfo{year}{2000}).

\bibitem[{\citenamefont{Eto and Nazarov}(2000)}]{eto}
\bibinfo{author}{\bibfnamefont{M.}~\bibnamefont{Eto}} \bibnamefont{and}
  \bibinfo{author}{\bibfnamefont{Y.~V.} \bibnamefont{Nazarov}},
  \bibinfo{journal}{Phys. Rev. Lett.} \textbf{\bibinfo{volume}{85}},
  \bibinfo{pages}{1306} (\bibinfo{year}{2000}).

\bibitem[{\citenamefont{Pustilnik and Glazman}(2000)}]{pustilnik2}
\bibinfo{author}{\bibfnamefont{M.}~\bibnamefont{Pustilnik}} \bibnamefont{and}
  \bibinfo{author}{\bibfnamefont{L.~I.} \bibnamefont{Glazman}},
  \bibinfo{journal}{Phys. Rev. Lett.} \textbf{\bibinfo{volume}{85}},
  \bibinfo{pages}{2993} (\bibinfo{year}{2000}).

\bibitem[{\citenamefont{Pustilnik and
  Glazman}(2001{\natexlab{a}})}]{pustilnik3}
\bibinfo{author}{\bibfnamefont{M.}~\bibnamefont{Pustilnik}} \bibnamefont{and}
  \bibinfo{author}{\bibfnamefont{L.~I.} \bibnamefont{Glazman}},
  \bibinfo{journal}{Phys. Rev. B} \textbf{\bibinfo{volume}{64}},
  \bibinfo{pages}{045328} (\bibinfo{year}{2001}{\natexlab{a}}).

\bibitem[{\citenamefont{Hofstetter and Zarand}(2004)}]{2stage-dot}
\bibinfo{author}{\bibfnamefont{W.}~\bibnamefont{Hofstetter}} \bibnamefont{and}
  \bibinfo{author}{\bibfnamefont{G.}~\bibnamefont{Zarand}},
  \bibinfo{journal}{Phys. Rev. B} \textbf{\bibinfo{volume}{69}},
  \bibinfo{pages}{235301} (\bibinfo{year}{2004}).

\bibitem[{\citenamefont{Pustilnik and
  Glazman}(2001{\natexlab{b}})}]{pustilnik4}
\bibinfo{author}{\bibfnamefont{M.}~\bibnamefont{Pustilnik}} \bibnamefont{and}
  \bibinfo{author}{\bibfnamefont{L.~I.} \bibnamefont{Glazman}},
  \bibinfo{journal}{Phys. Rev. Lett.} \textbf{\bibinfo{volume}{87}},
  \bibinfo{pages}{216601} (\bibinfo{year}{2001}{\natexlab{b}}).

\bibitem[{\citenamefont{Hofstetter and Schoeller}(2001)}]{hofstetter}
\bibinfo{author}{\bibfnamefont{W.}~\bibnamefont{Hofstetter}} \bibnamefont{and}
  \bibinfo{author}{\bibfnamefont{H.}~\bibnamefont{Schoeller}},
  \bibinfo{journal}{Phys. Rev. Lett.} \textbf{\bibinfo{volume}{88}},
  \bibinfo{pages}{016803} (\bibinfo{year}{2001}).

\bibitem[{\citenamefont{Lucignano et~al.}(2004)\citenamefont{Lucignano,
  Jouault, and Tagliacozzo}}]{procolo2}
\bibinfo{author}{\bibfnamefont{P.}~\bibnamefont{Lucignano}},
  \bibinfo{author}{\bibfnamefont{B.}~\bibnamefont{Jouault}}, \bibnamefont{and}
  \bibinfo{author}{\bibfnamefont{A.}~\bibnamefont{Tagliacozzo}},
  \bibinfo{journal}{Phys. Rev. B} \textbf{\bibinfo{volume}{69}},
  \bibinfo{pages}{045314} (\bibinfo{year}{2004}).

\bibitem[{\citenamefont{Lucignano et~al.}(2005)\citenamefont{Lucignano,
  Jouault, Tagliacozzo, and Altshuler}}]{procolo}
\bibinfo{author}{\bibfnamefont{P.}~\bibnamefont{Lucignano}},
  \bibinfo{author}{\bibfnamefont{B.}~\bibnamefont{Jouault}},
  \bibinfo{author}{\bibfnamefont{A.}~\bibnamefont{Tagliacozzo}},
  \bibnamefont{and} \bibinfo{author}{\bibfnamefont{B.~L.}
  \bibnamefont{Altshuler}}, \bibinfo{journal}{Phys. Rev. B}
  \textbf{\bibinfo{volume}{71}}, \bibinfo{pages}{121310}
  (\bibinfo{year}{2005}).

\bibitem[{\citenamefont{Khaetskii and Nazarov}(2000)}]{khaetskii_nazarov}
\bibinfo{author}{\bibfnamefont{A.~V.} \bibnamefont{Khaetskii}}
  \bibnamefont{and} \bibinfo{author}{\bibfnamefont{Y.~V.}
  \bibnamefont{Nazarov}}, \bibinfo{journal}{Phys. Rev. B}
  \textbf{\bibinfo{volume}{61}}, \bibinfo{pages}{12639} (\bibinfo{year}{2000}).

\bibitem[{\citenamefont{Stepanenko and Bonesteel}(2004)}]{SOI-qubit-1}
\bibinfo{author}{\bibfnamefont{D.}~\bibnamefont{Stepanenko}} \bibnamefont{and}
  \bibinfo{author}{\bibfnamefont{N.~E.} \bibnamefont{Bonesteel}},
  \bibinfo{journal}{Phys. Rev. Lett.} \textbf{\bibinfo{volume}{93}},
  \bibinfo{pages}{140501} (\bibinfo{year}{2004}).

\bibitem[{\citenamefont{Debald and Emary}(2005)}]{SOI-qubit-2}
\bibinfo{author}{\bibfnamefont{S.}~\bibnamefont{Debald}} \bibnamefont{and}
  \bibinfo{author}{\bibfnamefont{C.}~\bibnamefont{Emary}},
  \bibinfo{journal}{Phys. Rev. Lett.} \textbf{\bibinfo{volume}{94}},
  \bibinfo{pages}{226803} (\bibinfo{year}{2005}).

\bibitem[{\citenamefont{Golovach et~al.}(2006)\citenamefont{Golovach, Borhani,
  and Loss}}]{SOI-qubit-3}
\bibinfo{author}{\bibfnamefont{V.~N.} \bibnamefont{Golovach}},
  \bibinfo{author}{\bibfnamefont{M.}~\bibnamefont{Borhani}}, \bibnamefont{and}
  \bibinfo{author}{\bibfnamefont{D.}~\bibnamefont{Loss}},
  \bibinfo{journal}{Phys. Rev. B} \textbf{\bibinfo{volume}{74}},
  \bibinfo{pages}{165319} (\bibinfo{year}{2006}).

\bibitem[{\citenamefont{Baruffa et~al.}(2010)\citenamefont{Baruffa, Stano, and
  Fabian}}]{SOI-two-electrons}
\bibinfo{author}{\bibfnamefont{F.}~\bibnamefont{Baruffa}},
  \bibinfo{author}{\bibfnamefont{P.}~\bibnamefont{Stano}}, \bibnamefont{and}
  \bibinfo{author}{\bibfnamefont{J.}~\bibnamefont{Fabian}},
  \bibinfo{journal}{Phys. Rev. Lett.} \textbf{\bibinfo{volume}{104}},
  \bibinfo{pages}{126401} (\bibinfo{year}{2010}).

\bibitem[{\citenamefont{Bulla et~al.}(2008)\citenamefont{Bulla, Costi, and
  Pruschke}}]{bullaRMP}
\bibinfo{author}{\bibfnamefont{R.}~\bibnamefont{Bulla}},
  \bibinfo{author}{\bibfnamefont{T.~A.} \bibnamefont{Costi}}, \bibnamefont{and}
  \bibinfo{author}{\bibfnamefont{T.}~\bibnamefont{Pruschke}},
  \bibinfo{journal}{Rev. Mod. Phys.} \textbf{\bibinfo{volume}{80}},
  \bibinfo{pages}{395} (\bibinfo{year}{2008}).

\bibitem[{\citenamefont{Lucignano et~al.}(2007)\citenamefont{Lucignano,
  Stefanski, Tagliacozzo, and Bulka}}]{procolo3}
\bibinfo{author}{\bibfnamefont{P.}~\bibnamefont{Lucignano}},
  \bibinfo{author}{\bibfnamefont{P.}~\bibnamefont{Stefanski}},
  \bibinfo{author}{\bibfnamefont{A.}~\bibnamefont{Tagliacozzo}},
  \bibnamefont{and} \bibinfo{author}{\bibfnamefont{B.~R.} \bibnamefont{Bulka}},
  \bibinfo{journal}{Curr. Appl. Phys.}  (\bibinfo{year}{2007}).

\bibitem[{\citenamefont{Bychkov and Rashba}(1984)}]{rashba}
\bibinfo{author}{\bibfnamefont{Y.~A.} \bibnamefont{Bychkov}} \bibnamefont{and}
  \bibinfo{author}{\bibfnamefont{E.~I.} \bibnamefont{Rashba}},
  \bibinfo{journal}{J. Phys. C} \textbf{\bibinfo{volume}{17}},
  \bibinfo{pages}{6039} (\bibinfo{year}{1984}).

\bibitem[{\citenamefont{Dresselhaus}(1955)}]{dresselhaus}
\bibinfo{author}{\bibfnamefont{G.}~\bibnamefont{Dresselhaus}},
  \bibinfo{journal}{Phys. Rev.} \textbf{\bibinfo{volume}{100}},
  \bibinfo{pages}{580} (\bibinfo{year}{1955}).

\bibitem[{\citenamefont{Lucignano et~al.}(2010)\citenamefont{Lucignano,
  Fabrizio, and Tagliacozzo}}]{Lucignano_physica_E}
\bibinfo{author}{\bibfnamefont{P.}~\bibnamefont{Lucignano}},
  \bibinfo{author}{\bibfnamefont{M.}~\bibnamefont{Fabrizio}}, \bibnamefont{and}
  \bibinfo{author}{\bibfnamefont{A.}~\bibnamefont{Tagliacozzo}},
  \bibinfo{journal}{Physica E: Low-dimensional Systems and Nanostructures}
  \textbf{\bibinfo{volume}{42}}, \bibinfo{pages}{860 } (\bibinfo{year}{2010}),
  ISSN \bibinfo{issn}{1386-9477}.

\bibitem[{\citenamefont{Vojta et~al.}(2002)\citenamefont{Vojta, Bulla, and
  Hofstetter}}]{refereea1}
\bibinfo{author}{\bibfnamefont{M.}~\bibnamefont{Vojta}},
  \bibinfo{author}{\bibfnamefont{R.}~\bibnamefont{Bulla}}, \bibnamefont{and}
  \bibinfo{author}{\bibfnamefont{W.}~\bibnamefont{Hofstetter}},
  \bibinfo{journal}{Phys. Rev. B} \textbf{\bibinfo{volume}{65}},
  \bibinfo{pages}{140405} (\bibinfo{year}{2002}).

\bibitem[{\citenamefont{Cornaglia and Grempel}(2005)}]{refereea2}
\bibinfo{author}{\bibfnamefont{P.~S.} \bibnamefont{Cornaglia}}
  \bibnamefont{and} \bibinfo{author}{\bibfnamefont{D.~R.}
  \bibnamefont{Grempel}}, \bibinfo{journal}{Phys. Rev. B}
  \textbf{\bibinfo{volume}{71}}, \bibinfo{pages}{075305}
  (\bibinfo{year}{2005}).

\bibitem[{\citenamefont{Koller et~al.}(2005)\citenamefont{Koller, Hewson, and
  Meyer}}]{Hewson-underscreened}
\bibinfo{author}{\bibfnamefont{W.}~\bibnamefont{Koller}},
  \bibinfo{author}{\bibfnamefont{A.~C.} \bibnamefont{Hewson}},
  \bibnamefont{and} \bibinfo{author}{\bibfnamefont{D.}~\bibnamefont{Meyer}},
  \bibinfo{journal}{Phys. Rev. B} \textbf{\bibinfo{volume}{72}},
  \bibinfo{pages}{045117} (\bibinfo{year}{2005}).

\bibitem[{\citenamefont{Nozi{\`e}res}(1974)}]{Nozieres-JLTP}
\bibinfo{author}{\bibfnamefont{P.}~\bibnamefont{Nozi{\`e}res}},
  \bibinfo{journal}{J. Low Temp. Phys.} \textbf{\bibinfo{volume}{17}},
  \bibinfo{pages}{31} (\bibinfo{year}{1974}).

\bibitem[{\citenamefont{Hewson}(1997)}]{Hewson}
\bibinfo{author}{\bibfnamefont{A.}~\bibnamefont{Hewson}},
  \emph{\bibinfo{title}{The Kondo Problem to Heavy Fermions}}
  (\bibinfo{publisher}{Cambridge University Press}, \bibinfo{year}{1997}).

\bibitem[{\citenamefont{\v{Z}itko et~al.}(2008)\citenamefont{\v{Z}itko, Peters,
  and Pruschke}}]{Rock1}
\bibinfo{author}{\bibfnamefont{R.}~\bibnamefont{\v{Z}itko}},
  \bibinfo{author}{\bibfnamefont{R.}~\bibnamefont{Peters}}, \bibnamefont{and}
  \bibinfo{author}{\bibfnamefont{T.}~\bibnamefont{Pruschke}},
  \bibinfo{journal}{Phys. Rev. B} \textbf{\bibinfo{volume}{78}},
  \bibinfo{pages}{224404} (\bibinfo{year}{2008}).

\bibitem[{\citenamefont{\v{Z}itko et~al.}(2009)\citenamefont{\v{Z}itko, Peters,
  and Pruschke}}]{Rock2}
\bibinfo{author}{\bibfnamefont{R.}~\bibnamefont{\v{Z}itko}},
  \bibinfo{author}{\bibfnamefont{R.}~\bibnamefont{Peters}}, \bibnamefont{and}
  \bibinfo{author}{\bibfnamefont{T.}~\bibnamefont{Pruschke}},
  \bibinfo{journal}{New Journal of Physics} \textbf{\bibinfo{volume}{11}},
  \bibinfo{pages}{053003} (\bibinfo{year}{2009}).

\bibitem[{\citenamefont{Costi}(2000)}]{Costi-magnetic-field}
\bibinfo{author}{\bibfnamefont{T.~A.} \bibnamefont{Costi}},
  \bibinfo{journal}{Phys. Rev. Lett.} \textbf{\bibinfo{volume}{85}},
  \bibinfo{pages}{1504} (\bibinfo{year}{2000}).

\bibitem[{\citenamefont{Roch et~al.}(2009)\citenamefont{Roch, Florens, Costi,
  Wernsdorfer, and Balestro}}]{Costi-magnetic-field-underscreened}
\bibinfo{author}{\bibfnamefont{N.}~\bibnamefont{Roch}},
  \bibinfo{author}{\bibfnamefont{S.}~\bibnamefont{Florens}},
  \bibinfo{author}{\bibfnamefont{T.~A.} \bibnamefont{Costi}},
  \bibinfo{author}{\bibfnamefont{W.}~\bibnamefont{Wernsdorfer}},
  \bibnamefont{and} \bibinfo{author}{\bibfnamefont{F.}~\bibnamefont{Balestro}},
  \bibinfo{journal}{Phys. Rev. Lett.} \textbf{\bibinfo{volume}{103}},
  \bibinfo{pages}{197202} (\bibinfo{year}{2009}).

\bibitem[{\citenamefont{Parks et~al.}(2010)\citenamefont{Parks, Champagne,
  Costi, Shum, Pasupathy, Neuscamman, Flores-Torres, Cornaglia, Aligia,
  Balseiro et~al.}}]{Parks2010}
\bibinfo{author}{\bibfnamefont{J.~J.} \bibnamefont{Parks}},
  \bibinfo{author}{\bibfnamefont{A.~R.} \bibnamefont{Champagne}},
  \bibinfo{author}{\bibfnamefont{T.~A.} \bibnamefont{Costi}},
  \bibinfo{author}{\bibfnamefont{W.~W.} \bibnamefont{Shum}},
  \bibinfo{author}{\bibfnamefont{A.~N.} \bibnamefont{Pasupathy}},
  \bibinfo{author}{\bibfnamefont{E.}~\bibnamefont{Neuscamman}},
  \bibinfo{author}{\bibfnamefont{S.}~\bibnamefont{Flores-Torres}},
  \bibinfo{author}{\bibfnamefont{P.~S.} \bibnamefont{Cornaglia}},
  \bibinfo{author}{\bibfnamefont{A.~A.} \bibnamefont{Aligia}},
  \bibinfo{author}{\bibfnamefont{C.~A.} \bibnamefont{Balseiro}},
  \bibnamefont{et~al.}, \bibinfo{journal}{Science}
  \textbf{\bibinfo{volume}{328}}, \bibinfo{pages}{1370} (\bibinfo{year}{2010}).

\end{thebibliography}

\end{document}